\documentstyle[proceedings,psfig]{crckapb}
\begin{opening}
\title{THE AGE OF THE GALACTIC DISK}
\author{Giovanni CARRARO}
\institute{Department of Astronomy, Padova University\\
Vicolo Osservatorio  5, I-35122, Padova, Italy}
\end{opening}
\runningtitle{The age of the Galactic Disk}
\begin{document}

\begin{abstract}
I review different methods devised to derive the age 
of the Galactic Disk, namely the Radio-active Decay (RD), the
Cool White Dwarfs Luminosity Function (CWDLF), old open clusters (OOC)
and the Color Magnitude Diagram (CMD)
of the stars in the solar vicinity. I argue that the disk is likely
to be 8-10 Gyr old. Since the bulk of globulars has an age around 13 Gyrs,
the possibility emerges that the Galaxy experienced a minimum of Star
Formation at the end of the halo/bulge formation. This minimum
might reflect the time at which the Galaxy started to acquire material
to form the disk inside-out.
\end{abstract}

\section{Introduction}
Disks are quite common structures in the Universe. 
According to the classical theory of White \& Rees (1978)
spiral galaxies
are considered the seeds of galaxies assembly due to gas cooling 
inside spinning dark matter halos.
Apart from spirals, disks are seen also 
in the center of giant ellipticals; damped Lyman $\alpha$ clouds are
commonly believed to be gaseous disks. Finally, S0 galaxies 
are expected to be disk galaxies depauperated of their gas.\\
However galaxian disks are fragile structures. 
N-body simulations convincingly
show that the merging of two equal mass galaxies is fatal for disks. 
They are simply destroyed. 
The observational counterpart is represented -just to mention a source-
in the IRAS database, which shows many disk 
on the way to be destroyed by major mergers.

\begin{figure*}
\centerline{\psfig{file=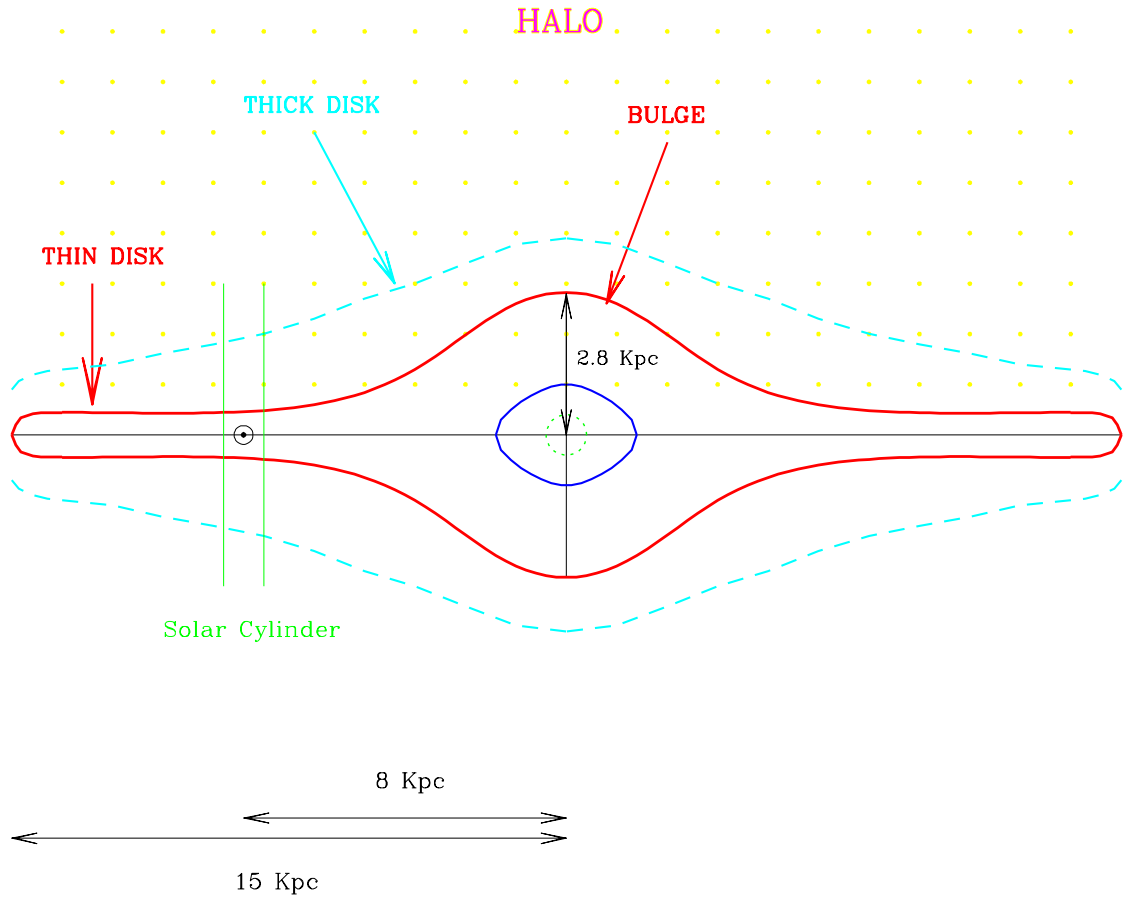,width=16cm}}
\caption{A sketch showing the vertical structure of the Galaxy. The solar
vicinity is shown as a cylinder with a radius of 1 kpc.}
\end{figure*}

Since our disk appears almost undamaged, this means that it did not suffer
from strong mergers since its formation, and that its past life was relatively
quiet. Of course we cannot forget that our disk has a warp, and that presently
a dwarf galaxy - Sagittarius - is going to merge with our Galaxy.\\
Just how the past life was quiet can be gauged by the ability of small 
galaxies to heat or thicken the disk.
Quinn et al (1993) pointed out that even satellites with masses of order 
$5\%-10\%$ of the disk mass can thicken the disk by a factor of two or more
if they dump all of their available orbital kinetic energy into the disk.
Hence the presence of disks with scale heights less than a few hundred 
parsecs ({\it thin disks})  
implies the absence of even minor encounters over the age of the 
{\it thin disk}. Nonetheless
it may be possible for disks formed early in the lives of spirals
to have been heated by satellites forming a thick disk within which a new 
{\it thin disk} was allowed to form. If this was the case, an estimate
of the {\it thin disk} age can fix the time at which the last encounter
occured.\\
The acute fragility of disks is a very important constraint on
the evolution of the environment of protogalaxies that develop into current
spirals. Either they were born in isolation (low density environment)
or they managed to remove all potentially disk-disturbing debris early 
enough that a normal disk had time to form.
It is clearly very interesting to look from our favorite observational point,
the solar system, at the disk population to try to establish its age.
This is important to constrain the zero point of chemical evolution
models, the relationship between the disk and the other galaxy components,
and to fix a rough chronology for the disk development and origin.

\section{Age indicators}
In the following I shall discuss five different methods
devised to obtain an estimate of the age of the galactic disk,
discussing their feasibility, robustness and limitations.\\
I shall start making a crucial point.
Most methods to infer the age of the galactic disk pretend, or try,
to give an age estimate for the entire disk
(or even for the Galaxy as a whole), by using 
age indicators located in the near solar vicinity (see Fig.~1).
Whether the solar neighborhood is really representative of all
the disk is an open question, which does not hold only
for this particular issue - the age of the disk -, 
but more generally for determining
the global chemical evolution and the SF history of the galactic disk 
looking at nearby indicators (Carraro et al 1998).
At present only old open clusters (although the sample is rather poor)
can be used to derive an estimate for the age of a
significant portion
of the galactic disk.
Moreover all the methods devised to address this topic make us of indicators 
located well inside the {\it thin disk}. Therefore I am going to discuss the
age of the galactic {\it thin disk} (Gilmore et al 1989).

\begin{figure*}
\centerline{\psfig{file=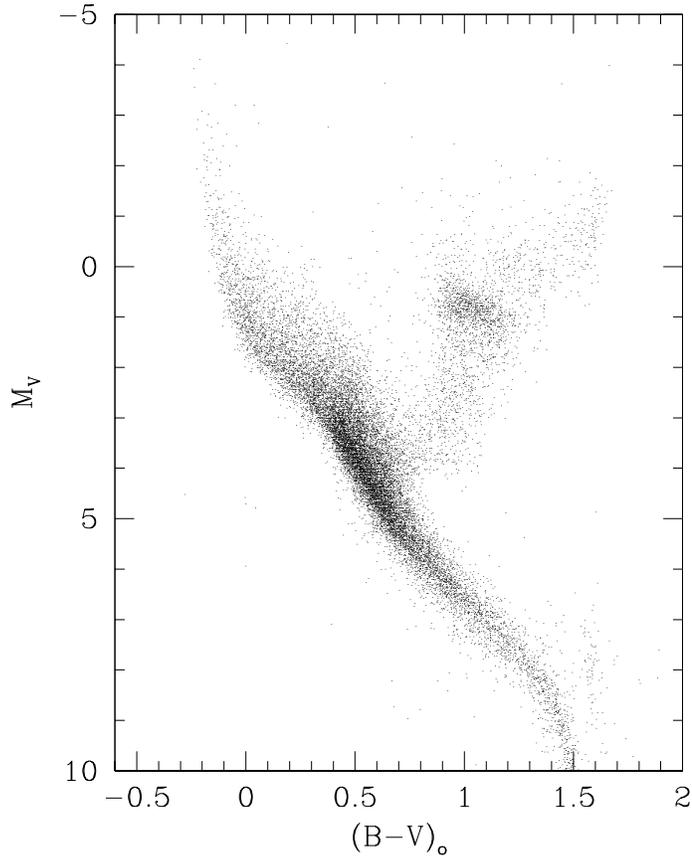,width=12cm}}
\caption{The CMD of nearby stars from Hypparcos. The red envelope of the subgiant region
is used to set a minimum age of the galactic disc.}
\end{figure*}

\subsection{Cool White Dwarfs Luminosity Function}
The idea that the coolest white dwarfs (WD) can be used to determine the 
age of the local galactic disk dates back to Schmidt (1959).
His idea was the following: if we find the coolest white dwarf (just looking
at the downturn of their luminosity function) 
and measure their cooling time,
we can find the age of the disk just adding to this time the lifetime
of the WD progenitor. Moreover WDs are easy to be modelled, much easier than 
Main Sequence (MS) stars (Mestel 1952).
At that time it was not possible to detect the turn-down, and the idea failed.
Moreover some uncertainties in the models arose, in particular the crystallization.
In detail, at decreasing temperature ions start to crystallize, and the energy 
necessary to maintain the ions lattice causes the WD to cool to invisibility
very rapidly.\\
More recently Winget et al (1987) 
succeeded to find the turn-down, and since then many authors
built up CWDLF (Leggett et al 1998).
In addition theorists now agree that the onset of rapid cooling is 
not reached for the dominant
stellar mass in the WD population ($0.6 M_{\odot}$) until well below the luminosity
of the observed shortfall.
The most recent work is from Knox et al (1999). They find 52 CWDs 
using common proper motion binaries. Their sample is the only 
complete one both in
luminosity and in proper motion. The resulting LF 
(see their Figs.~19,20 and 22) shows a clear short-fall at 
$log L/L_{\odot}~=~-4.2$, 
and predicts a larger number of WDs per unit volume, as compared with
previous determination of the LF.
Knox et al (1999) compared their LF with two sets of stellar models, 
by Wood  (1992) and Garcia-Berro et al (1997).
The best fit is obtained using Wood (1992) models, and provides a disk age
of $9.0\pm1.0$ Gyr. Garcia-Berro et al (1997) models do not fit the LF maximum, 
and provide a somewhat greater age.

\subsection{The Hypparcos CMD}
The CMD provided by Hipparcos satellite
includes stars within about 150 parsecs from the Sun
(see Fig.~2). 
It shows a well defined
MS, a prominent clump of He-burning stars and a subgiant region.
It represents of course a mix of stellar populations with spreads in age and 
in metallicity. 
The dating features used by Jimenez et al (1998) is the red envelope
of the sub-giant region, and the methods adopted is the isochrone fitting one.
Since the metallicity of the stars is not known - say from
observations - the authors assume that the spread in color of the clump
is a good indicator of a spread in metallicity. This is a rather crude 
approximation, as discussed by Girardi (this meeting).
Nevertheless, it is possible to obtain a minimum age for this sample,
assuming that the star populating the lower red envelope 
of the subgiant region has the maximum metallicity, since this metallicity
provides the minimum age.
So doing they obtain a minimum age for the disk of $8$ Gyr.

\subsection{Nearby $F$ \& $G$ stars}
Another method - although rather difficult - 
is the direct age estimate of single stars, like the 187
stars sample of Edvardsson et al (1993).
To get an age estimate one needs star photometry, spectroscopy
and distance, to put them in the $M_{V}-logT_{e}$ plane. In the case of the
Edvardsson et al sample ages are inferred on the Vandenberg scale (1985).
Removing from the sample the presumed thick disk stars, or stars whose orbital
motions are not that of thin disk stars, a reasonable estimate for the age
of the local galactic disk is $9-11$ Gyr.
Recently the Edvardsson et al sample has been revised by Ng  \& Bertelli (1997),
who re-computed the ages of those stars taking into account new distances from
Hipparcos, correcting for the Lutz-Kelker effect, and putting them
in the Bertelli et al (1994) scale. 
At older ages the new ages are slightly older, but 
the conclusion on the disk age is roughly the same.

\subsection{The Radio-active Clock}
Butcher (1987)  proposed to derive the age of the Galaxy 
by observing the radio-active nuclide $^{232}Th$ in stars of different ages, 
and relating the nucleosynthesis timescale to the stellar and galactic evolution.
He considered the evolution of Nd, a stable nuclide
and Th (half life 14 Gyr).  
The first point to stress is the extreme 
weakness of the spectral features in the
measured stars spectra: in particular the Th line falls in a blend with Co.
The errors related to the derived abundance are around 0.1 dex.
The idea underlying this method is that after its formation,
a stars does not modify its envelope abundances of Th and Nd
but for radio-active decay.
By measuring the ratio of the abundances of these elements,
$[Th/Nd]$ in stars of different age, it is possible 
to reconstruct the decay evolution of Th. 
The basic assumption is that the growth rate of the two elements
is the same, although Th is a r-process and Nd is partly a r- and 
partly a s-process. 
So doing, they conclude that no reliable chemical evolution model
can account for this distribution without assuming an age less than 
$9$ Gyr. The same result has been obtained by Morell et al (1992)
who made a new abundance analysis on the same sample.\\

Clayton (1988) criticized these results, stressing that although 
the precise nature of r- and s- process is not clear enough,
in principle one has to take into account their different  evolution.
In particular assuming that the contribution to the
Nd abundance is about half from r-
and half from s-processes,
he showed that a simple model of chemical evolution 
can account for the Butcher distribution 
assuming ages greater than 12 Gyr,
and concluding: "{\it An unbiased look at all methods together favours
an age greater than 12 Gyr, although no single method is reliable. I point
out that each nuclear method is still amenable to further improvements,
but they alone will not be able to determine the Galaxy's age. Only a 
detailed and specific and correct model for the growth and chemical evolution
of the solar neighbourhood can enable the galactic age to be inferred from
radioactivity}.

\begin{figure*}
\centerline{\psfig{file=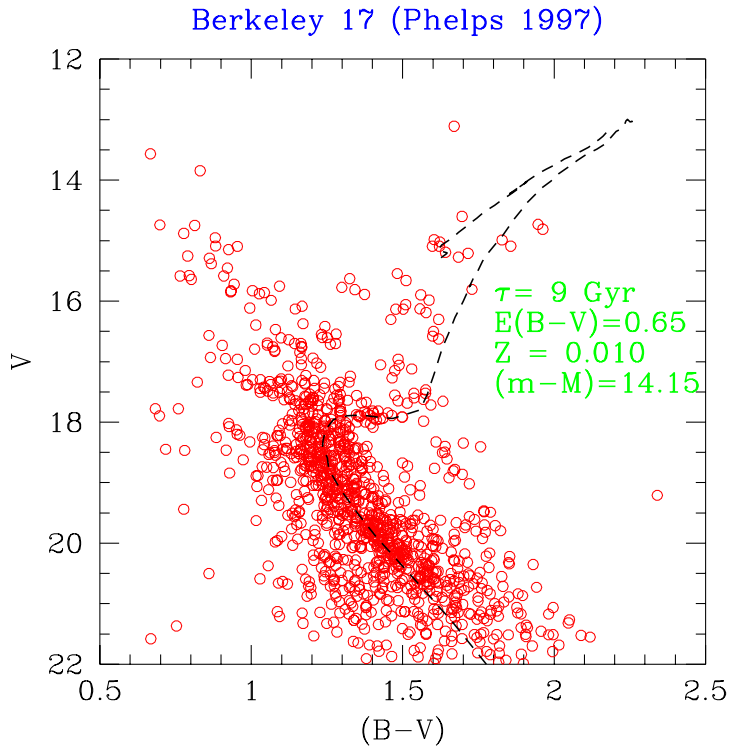,width=12cm}}
\caption{The CMD of Berkeley~17 from the V,I 
photometry of Phelps (1997). Overimposed is an isochrone 
from Girardi et al 1999 for the parameters
listed in the plot.}
\end{figure*}

\subsection{Old Open Clusters}
Old open cluster are well suited to address 
many issues concerning our disk (Friel 1995, Carraro et al 1998). 
For the present topic, they are in principle
more suitable than the other indicators
to derive a lower limit for the age of the galactic disk. 
In fact they are distributed in 
a larger portion of the disk. 
Good data have been obtained recently for clusters
older than M~67, so it is actually feasible to use this sample to determine
the age of the disk.
However there is still debate on the role of the oldest open cluster.
NGC 6791, often quoted as the oldest cluster (8-9 Gyr,
Carraro et al 1999c), is a rather special object.  
Its nature is not completely clear, and somebody is suggesting that it could be 
a bulge globular pushed away by the bar, or the core of a dwarf
spheroidal tidally stripped by our Galaxy (see Carraro et al 1999a
for a detailed analysis on this cluster).
Recently another cluster came out to be very old, Berkeley~17
(Fig.~3).  
It is actually quite old, with an age around 9 Gyr (Carraro et al 1999b), 
although optical photometry is not very good yet, and 
should be improved, being this cluster so important.
If Berkeley~17 marks the age of the disk, 
its minimum age is around 9 Gyr.\\
However the main drawback of open clusters is that their average lifetime
is of the order of some $10^8$ yr (Grenon 1990), so many old clusters might have 
been destroyed. 
Therefore the statistics of old open clusters is rather poor,
and in principle they can provide only a lower limit for the age
of the thin disk (Carraro et al 1999c). Anyhow their dating is rather 
simple and robust.

\begin{figure*}
\centerline{\psfig{file=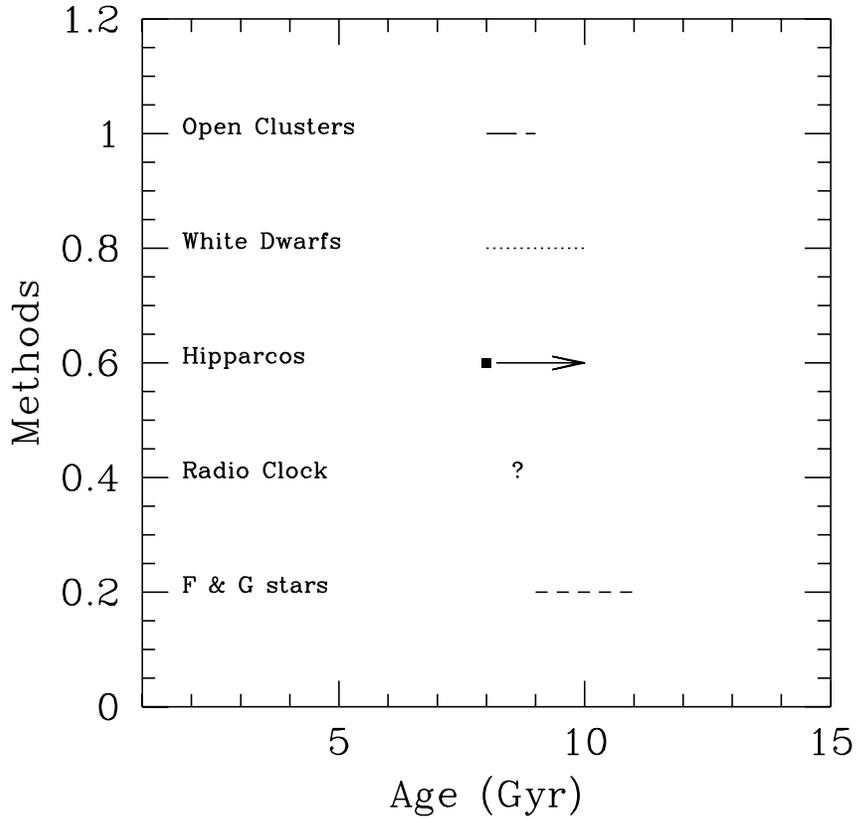,width=12cm}}
\caption{Summary of disk ages from different methods.}
\end{figure*}

\section{Conclusions}
In the past several reviews were dedicated to the disk age issue. I would like
to remind the reviews from Sandage (1990), van den Bergh (1990)
and Grenon (1990).\\
Sandage at the 1989 Kingston Conference in Canada
concluded that the disk overlaps in age the halo,
whereas van den Bergh, at the same meeting, suggested that the disk
is somewhat younger than the halo. \\
Obviously the question of a possible star formation
delay at the end of the halo assembly depends on the age of the halo,
and on the age of the disk. The age of the halo comes from the mean
globular clusters age ($12-15$ Gyr at that time), 
whereas the age of the disk comes from a variety
of methods, going from old open cluster to white dwarfs, from
radioactive decay to stars in the solar neighbourhood. 
At that time the oldest open cluster was NGC~6791, with an age around 12 Gyr
according to Sandage, who quoted Janes (1988), and of 7 Gyr according to van den Bergh,
who quoted a preliminary work by Demarque et al (1992).\\
The conclusion of van den Bergh is supported also by Grenon (1990), who
in addition stressed that in the solar vicinity there is a group of 
metal rich stars which seem to be older than open clusters, but whose 
birthplaces might be inside the bulge.\\

\noindent
Ten years after the situation is not much different.\\
Gratton et al (1997) reported a mean age of the bulk of the halo globulars
around $13$ Gyr, which holds for all the halo clusters. There is indeed
the evidence of a population of very young globulars (Pal~12 for instance),
whose belonging to the halo is controversial.\\
Summarizing all the data I discussed above (see also Fig.~4) a plausible age for
the disk is in the range $8-10$ Gyr. Note that the age scale in the case
of F \& G stars and open clusters is the same as for globulars.\\
This seems to suggest the occurence of a hiatus, or minimun
in the star formation history 
of the Galaxy, which might reflect the end of the halo/bulge formation.
Afterwards the Galaxy started to acquire material to form the disk in 
an inside-out scenario.\\

\noindent
{\bf Acknowledgements}\\
I thank L. Girardi and M. Grenon for useful discussions, and Francesca Matteucci
for her invitation.

\end{document}